\documentclass[11pt, onecolumn]{article}

\usepackage[english]{babel}
\usepackage{abstract}

\usepackage{amsmath}
\usepackage{amssymb}
\usepackage{euscript}
\usepackage{hyperref}

\usepackage{cyr}
\usepackage{epsfig}

\makeatletter
\renewcommand{\@oddhead}{\small Published in Geomagnetism and Aeronomy, 2012, Vol.52, No.8 \hfill}
\renewcommand{\@oddfoot}{\hfill ---~\thepage~---\hfill}
\makeatother

\begin{document}




\begin{center}
\Large Calculation of the Solar Activity Effect on the Production Rate of Cosmogenic Radiocarbon in Polar Ice

\vspace{0.5cm}
\large A.V. Nesterenok$^1$ and V.O. Naidenov
\vspace{0.5cm}

\normalsize Ioffe Physical-Technical Institute, Russian Academy of Sciences, Polytechnicheskaya St. 26, 
Saint~Petersburg, 194021 Russia

Saint Petersburg State Polytechnic University, Polytechnicheskaya St. 29, Saint Petersburg, 195251~Russia

$^1$e-mail: alex-n10@yandex.ru
\end{center}

 \begin{abstract} 
  \noindent
The propagation of cosmic rays in the Earth$^{\prime}$s atmosphere is simulated. Calculations of the omnidirectional differential flux of neutrons for different solar activity levels are presented. The solar activity effect on the production rate of cosmogenic radiocarbon by the nuclear-interacting and muon components of cosmic rays in polar ice is studied. It has been obtained that the $\rm ^{14}C$ production rate in ice by the cosmic ray nuclear-interacting component is lower or higher than the average value by 30\% during periods of solar activity maxima or minima, respectively. Calculations of the altitudinal dependence of the radiocarbon production rate in ice by the cosmic ray components are illustrated.
\end{abstract}

\noindent

\section{Introduction}

Cosmogenic radionuclides are produced in the Earth$^{\prime}$s atmosphere and in matter on the Earth$^{\prime}$s surface in nuclear reactions caused by primary and secondary cosmic rays. The data on the cosmogenic radionuclide concentration in natural archives, such as tree rings and soil and ice layers, bear information on changes in the primary cosmic ray intensity.

Radioactive isotope $\rm ^{14}C$ is accumulated in polar ice together with atmospheric air precipitation in ice at the stage of glacier ice formation from firn. The nuclide production in reactions caused by secondary cosmic rays (in situ production) is another mechanism by which radiocarbon is accumulated in ice. Spallation reactions of oxygen atoms by neutrons of the cosmic ray nuclear-interacting component mainly contribute to the radiocarbon production in the upper ice layers. At depths larger than the depth of penetration of the cosmic ray neutrons (several meters), $\rm ^{14}C$ is produced in reactions caused by cosmic ray muons.

The $\rm ^{14}C$ production rate in ice by secondary cosmic rays depends on the cosmic ray intensity. Geomagnetic field variations do not affect galactic cosmic ray fluxes in polar regions. Thus, data on the in situ $\rm ^{14}C$ concentration in polar ice samples can be used as chronicles of solar activity variations in the past \cite{Lal2005}. The present work studies the effect of solar activity variations on the radiocarbon production rate in ice by different cosmic ray components. The aim of this work is to quantitatively estimate the effect of the considered factor.

\noindent

\section{Numerical calculations}
\label{Nc}
\noindent

\subsection{Galactic Cosmic Ray Spectrum}

At energies higher than several hundreds of MeV, galactic cosmic rays make the main contribution to the particle flux in the near Earth space. Protons and alpha particles account for about 90 and 10\% of the total number of galactic cosmic ray particles, respectively. The particle fraction of heavier nuclei does not exceed 1\%. The flux of galactic cosmic rays has a high degree of isotropy \cite{Panasyuk2005}.

For particle energies lower than 10 GeV/nucleon, the flux of galactic cosmic rays in the near Earth space depends substantially on the solar activity level. The expression for the differential energy spectrum of \textit{i} particles with mass and charge numbers $A_i$ and $Z_i$ has the following form in the "force field" model \cite{Gleeson1968}:

\begin{equation}
	 	J_{i} (E) = J_{i, LIS} (E+\Phi_{i})\frac{E(E+2m_{p}c^{2})}{(E+\Phi_{i})(E+\Phi_{i}+2m_{p}c^{2})}
	 	\label{spectra1}
\end{equation}

\noindent
where $J_{i, LIS}(E)$ is the differential energy spectrum of the \textit{i} particles of galactic cosmic rays in the local interstellar space; $E$ is the particle kinetic energy per nucleon; $m_{p}c^{2}$ is the proton rest energy; $\Phi_{i}=(eZ_{i}/A_{i})\phi$; $e$ is the elementary charge; and $\phi$ is the parameter of cosmic ray modulation in the heliosphere.

The following expression from \cite{Usoskin2005} is used as a $J_{i, LIS}(E)$ approximation:

\begin{equation}
	J_{i, LIS}(E) = C_{i}\frac{p(E)^{-2.78}}{1+0.487p(E)^{-2.51}}
	\label{spectra2}
\end{equation}

\noindent
where parameter $E$ corresponds to the particle kinetic energy in GeV/nucleon, $C_{i}$ is the normalization factor, $C_{p} = 1.9 \times 10^4 \; \rm (m^2 \; s \; sr \; GeV)^{-1}$ for protons, and $C_{He}= 9.5\times 10^2 \; \rm (m^2 \; s \; sr \; GeV/nucleon)^{-1}$ for helium nuclei.

Usoskin et al. (2005) \cite{Usoskin2005} presented the results of a reconstruction of the solar modulation parameter based on neutron monitor data for 1951-2004. The monthly average values of modulation parameter $\phi$ vary from 0.3 to 2 GV, and the average $\phi$ value for the considered time interval is 0.69 GV. In our calculations, we selected the minimal and maximal values of the modulation parameter equal to 0.3 and 1.2 GV, respectively. A parameter value of 0.3 GV corresponds to the monthly average value of the modulation parameter at a minimum of cycle 22, and 1.2 GV corresponds to the average annual value of this parameter at the cycle maximum \cite{Usoskin2005}.

The form of the energy spectrum of galactic cosmic ray particles at low energies in the force field model depends on the charge-to-mass number ratio $Z/A$. For nuclei with charge numbers $Z \geq 2$, $Z/A \approx 1/2$. The contribution of nuclei with $Z \geq 2$ to cosmic ray particle fluxes in the Earth$^{\prime}$s matter was determined by multiplying the data obtained for $\rm ^{4}He$ nuclei of galactic cosmic rays by coefficient $k$, which is the ratio of the nucleon flux caused by the nuclei of elements with $Z \geq 2$ to such a flux caused by galactic cosmic ray $\rm ^{4}He$ nuclei. Based on the parameters of the energy spectra presented in \cite{WiebelSooth1998}, it was found that $k = 1.44$. 

We ignored the geomagnetic cutoff of the cosmic ray differential energy spectrum, since we consider the production of cosmogenic $\rm ^{14}C$ in polar ice at high geomagnetic latitudes.

\subsection{Atmospheric Model}

In the numerical model, we assumed that the Earth radius and the height of the atmosphere are 6371 and 100 km, respectively. It was also assumed that the atmosphere includes 0.755, 0.232, and 0.013 mass fractions of nitrogen, oxygen, and argon, respectively \cite{COESA1976}. The atmosphere was modeled in the form of concentric layers with a constant air density within a layer. It was assumed that the thickness of each atmospheric layer is 15 $\rm g/cm^{2}$. The total thickness of the atmosphere was taken to be equal to 1034 $\rm g/cm^2$. The dependence of the air density on altitude corresponds to the average air density at latitudes higher than $60^{\circ}$N according to the COSPAR data. In the calculations, we assumed that the Earth$^\prime$s surface layer is composed of $\rm H_{2}O$ ice with a density of 0.917 $\rm g/cm^{3}$.

\subsection{Physical Model}

The numerical code for modeling cascades of cosmic ray particles was written based on the GEANT4 9.4 simulation toolkit \cite{Agostinelli2003}. The processes of production, propagation, and interaction of baryons (neutrons, protons, short-lived baryons, and their antiparticles), leptons (electrons, positrons, and muons), mesons (pions and kaons), light nuclei, and gamma rays were taken into account in the calculations. The standard set of processes was used in modeling electromagnetic interactions \cite{Geant4}. Particles of the electron-photon component of secondary cosmic rays with energies lower than 10 MeV were not considered. Photonuclear and electronuclear processes, as well as decay processes, were taken into account in modeling. Low- and high-energy models based on the parametrization of experimental data were used to describe the processes of inelastic hadron scattering by nuclei. The Bertini intranuclear cascade model was used to model the processes of inelastic nucleon and meson scattering by nuclei at energies lower than 6 GeV. A model of the intranuclear binary cascade was used to describe inelastic interactions of light nuclei. The processes of capture of negative mesons and those of neutron capture and nuclear fission were taken into account in the calculations. To describe the interaction between neutrons and nuclei, we used the G4NDL 3.14 package of cross sections. This package includes data from the ENDF-B VI, JENDL, and other data libraries for neutron energies lower than 20 MeV and the JENDL/HE data for energies varying from 20 MeV to 3 GeV.

\subsection{Calculation of Particle Fluxes}

A flux of galactic cosmic ray particles with an energy higher than $E$ through an area of the upper atmospheric boundary is

\begin{equation*}
	F_{0i} = \int\limits_{2\pi} d\Omega \cos\theta \int\limits_{E}^{\infty} dE^{\prime} J_{0i}(E^{\prime})=\pi\int\limits_{E}^{\infty} dE^{\prime} J_{0i}(E^{\prime})
\end{equation*}

\noindent
where $J_{0i}(E)$ is the directed differential flux of galactic cosmic ray particles of sort $i$ in the near-Earth space and $\theta$ is the angle between the particle momentum vector and the direction toward the nadir point. The angular distribution of the number of particles that fall into the atmosphere through the area unit element satisfies the relationship $dF_{0i}/d\cos\theta \sim \cos\theta$.

The omnidirectional differential flux of $i$ particles with energy $E$ in a target at depth $z$ was calculated using the formula:

\begin{equation}
I_{i}^{\rm diff}(E,z) = \frac{F_{0p}}{N_{0p} \Delta E} \sum_{l} \frac{1}{\vert\cos\theta_{li}\vert} + k\frac{F_{0\alpha}}{N_{0\alpha}\Delta E} \sum_{l} \frac{1}{\vert\cos\theta_{li}\vert}
\label{flux1}
\end{equation}

\noindent
where the $p$ and $\alpha$ subscripts mean proton and alpha particle, respectively; $N_{0}$ is the number of particles of primary cosmic rays for which modeling was performed; the summation over $l$ means summation with respect to all $i$ particles that cross the specified $z$ level in a target and have energies in the interval $[E-\Delta E/2; E+\Delta E/2]$; $\theta_{li}$ is the zenith angle of the $i$th particle; and $k$ = 1.44. Particles with $\vert \cos \theta _{li} \vert < \rm 0.001$ for atmospheric layers and $\vert \cos \theta _{li} \vert < \rm 0.01$ for ice were ignored in sum (\ref{flux1}). The $1/\cos \theta$ coefficient under the sign of the sum is used to calculate the omnidirectional particle flux.

In each case, the number of primary particles, for which numerical simulation was performed, varied from 5 to 10 millions for protons and from 2 to 3 millions for alpha particles. The particle energy was chosen according to differential energy spectrum (\ref{spectra1}) and (\ref{spectra2}). Energies from 100 MeV to 1000 GeV for protons and from 100 MeV/nucleon to 1000 GeV/nucleon for alpha particles were considered. The calculations were performed at the Saint Petersburg branch of Joint Supercomputer Center of the Russian Academy of Sciences.

\section{Results of numerical simulations}

\subsection{Omnidirectional Differential Flux of Neutrons at Sea Level for the Conditions of Moderate Solar Activity}

\begin{figure}[t]
	\centering
	\includegraphics[width = 0.65\textwidth]{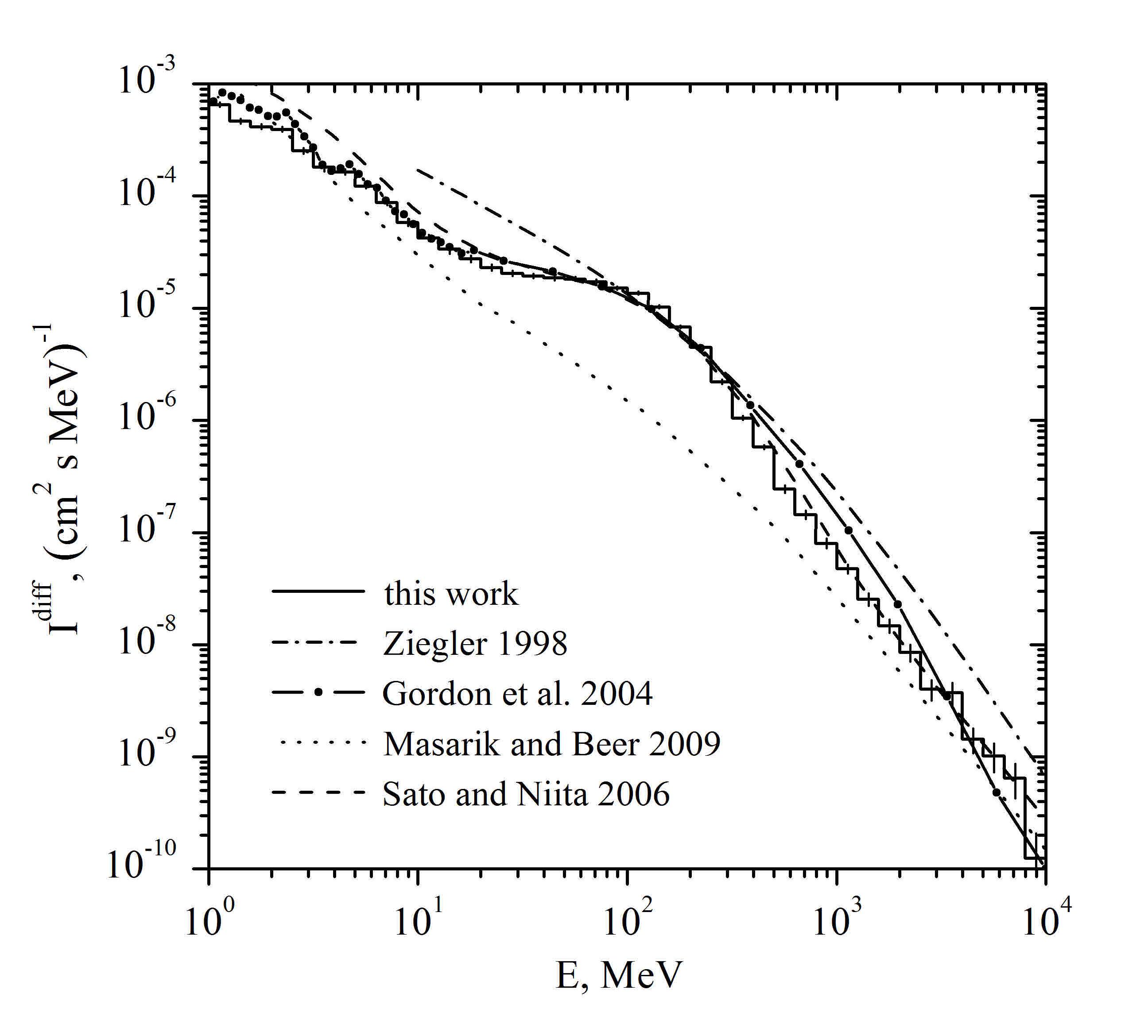}
		\caption{Our calculations of the omnidirectional differential neutron flux at sea level for the conditions of high geomagnetic latitudes and moderate solar activity (a stepped line). The data from \cite{Ziegler1998} are presented by a dot-and-dash line; the measurements performed by \cite{Gordon2004}, by a solid line with dots; the calculations made by \cite{Masarik2009}, by a dotted curve; the calculations performed by \cite{Sato2006}, by a dashed line.}
		\label{fig1}
\end{figure}

Figure \ref{fig1} illustrates calculations of the energy dependence of the omnidirectional differential neutron flux at sea level and moderate solar activity. Modulation parameter $\phi$ was taken equal to 0.69 GV in the calculations. For comparison, Fig. \ref{fig1} presents the energy dependences of the differential neutron flux according to the data of the studies \cite{Ziegler1998, Gordon2004, Masarik2009, Sato2006}.

The energy dependence of the differential neutron flux published in \cite{Ziegler1998} resulted from an approximation of the available differential flux measurements. The data correspond to the New York location. 

The differential flux of neutrons from \cite{Gordon2004} resulted from a normalization of the measurements in the Northern Hemisphere for high geomagnetic latitudes, moderate solar activity, and sea level according to \cite{Gordon2004}. The measurement error is caused by the error of cross sections of interactions between high-energy neutrons and the matter that were used to calculate the spectrometer response function. According to \cite{Gordon2004}, the error of the calculated response function values is about 10-15\% for neutron energies higher than 150 MeV and is lower for lower neutron energies, producing the measurement error of the same order of magnitude.

The omnidirectional differential neutron fluxes obtained in \cite{Masarik2009} resulted from simulations of cosmic ray propagation in the Earth$^\prime$s atmosphere. In \cite{Masarik2009}, numerical calculations were performed using the GEANT and MCNP program packages. In spite of the fact that we also used the GEANT program package in our calculations, the results of our modeling differ substantially from the data obtained in \cite{Masarik2009}. The reason of such discrepancy may be in different cross-section data used in both calculations. 

In \cite{Sato2006}, numerical calculations were performed using the PHITS program, JENDL/HE cross sections were used to model the interactions between nucleons and nuclei. The result from \cite{Sato2006} presented in Fig. \ref{fig1} corresponds to the calculations performed ignoring the effect of the Earth$^\prime$s surface. The results of our calculations are in good agreement with the data obtained in \cite{Sato2006} for neutron energies higher than 10 MeV.

At neutron energies 10-300 MeV, the calculations are in good agreement (within 30\%) with the measurements \cite{Gordon2004}. At neutron energies higher than 400 MeV, the difference between the calculated and experimental differential fluxes is substantial and reaches 200\%. A similar difference with the experiment is observed for the calculated differential neutron flux from \cite{Sato2006}. One of the possible causes of this difference consists in that the total JENDL/HE cross-sections used in both simulations are overstated at these energies. The difference in the cross-sections of the total interaction in 15\% can result in a twofold difference in particle fluxes on five to six free paths.

The differential flux of neutrons of secondary cosmic rays near the Earth$^\prime$s surface depends on the landscape and surface matter chemistry. A relatively small amount of hydrogen atoms in soil matter can substantially affect the neutron flux value. Disturbances in the neutron differential flux caused by water in soil decrease with increasing neutron energy: when the neutron energy is 10 MeV, the neutron flux disturbance is no more than 30\% \cite{Sato2006}. This phenomenon can be one of the causes of the differences between calculated differential fluxes and experimental data \cite{Gordon2004} at low energies (see Fig. \ref{fig1}).

\subsection{Omnidirectional Differential Flux of Neutrons
during Periods of Solar Activity Minima and Maxima}

Figure 2 illustrates the calculated energy dependence of the omnidirectional differential neutron flux for an atmospheric thickness of 660 $ \rm g/cm^2$, which corresponds to an altitude of about 3.5 km. The calculation results for cosmic ray modulation parameters of 0.3, 0.69, and 1.2 GV are presented. According to the calculations, the average ratio of the energy spectrum values for the periods of solar activity minima and maxima is about 1.6 for neutron energies reaching several hundred MeV and is smaller for higher energies. It is interesting that the ratio of the neutron monitor count rates during periods of solar activity minima and maxima is about 1.4 for high geomagnetic latitudes and lower atmospheric layers \cite{NM}. The difference between the calculated and neutron monitor data can be partially explained by the sensitivity of neutron monitors to the high-energy part of the neutron spectrum and to the cosmic ray muon component. Cascade processes can also result from interactions between atmospheric matter and galactic cosmic rays with higher energies than it follows from nuclear interaction models. As a result, calculations give a larger amplitude of variations in the secondary cosmic ray fluxes.

\begin{figure}[t]
	\centering
	\includegraphics[width = 0.65\textwidth]{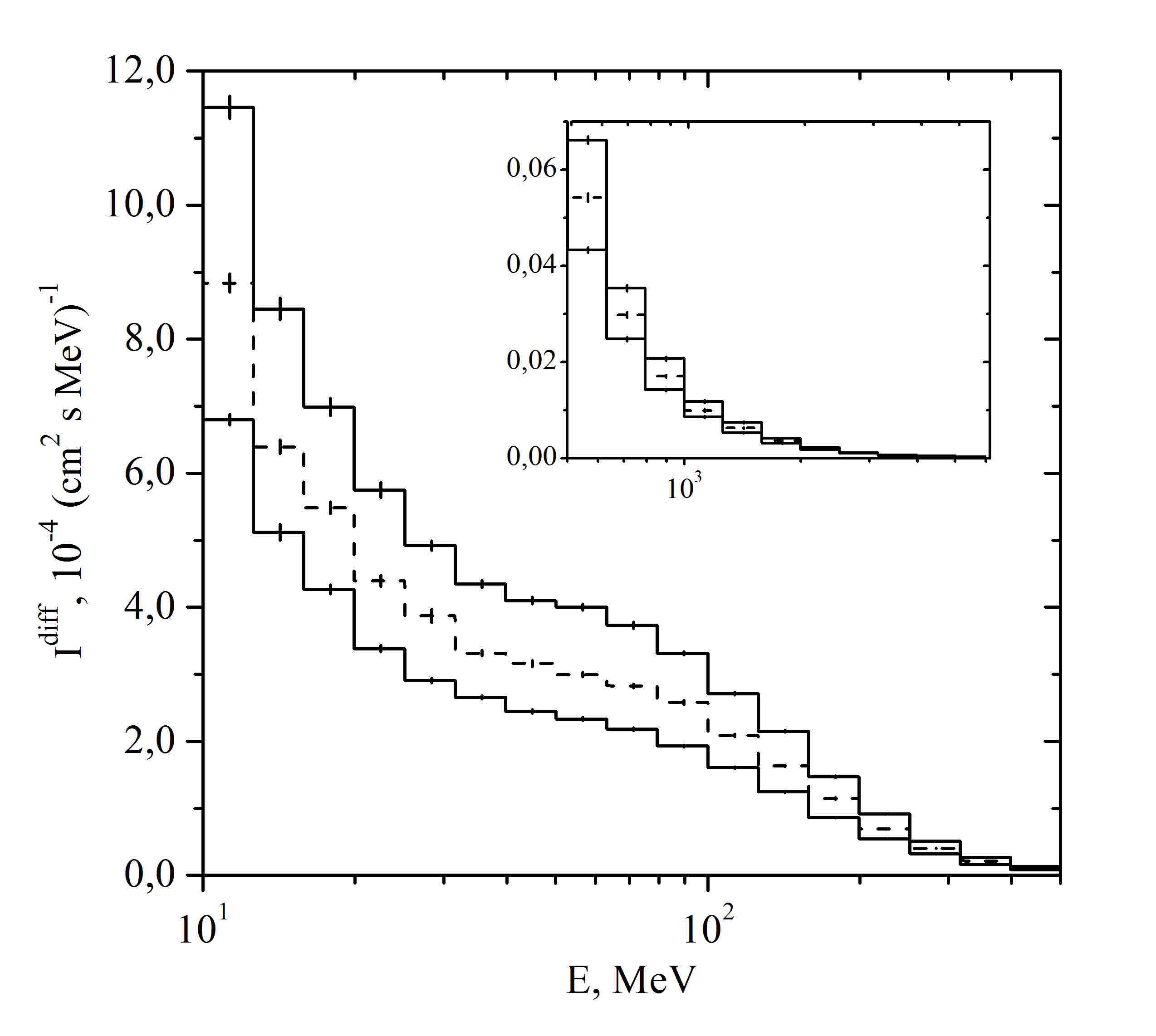}
		\caption{Calculations of the omnidirectional differential neutron flux for an atmospheric thickness of 660 $\rm g/cm^{2}$, for high geomagnetic latitudes, and three values of the solar modulation parameter: 0.3 (the upper curve), 0.69 and 1.2 GV (the lower curve).}
		\label{fig2}
\end{figure}

\section{Production of $\rm ^{14}C$ in ice by cosmic ray components}

\subsection{Production of $\rm ^{14}C$ in ice by nucleons of cosmic rays at high geomagnetic latitudes}

Reactions of oxygen nuclei spallation by neutrons make the main contribution (98\%) to the production of radiocarbon in ice by nucleons of the nuclear-interacting component of cosmic rays \cite{Nesterenok2012}. The results of the work \cite{Nesterenok2012} were used to calculate the average radiocarbon production rate in polar ice by the nuclear-interacting component of cosmic rays. The main part of radiocarbon (80-90\%) is produced in reactions induced by neutrons with energies 30-300 MeV. Based on calculations of the neutron differential flux in the atmosphere presented in the previous section, we can anticipate that the $\rm ^{14}C$ production rate in ice will be lower or higher than the average value by 30\% during periods of solar activity maxima or minima, respectively.

\subsection{Production of $\rm ^{14}C$ in ice by cosmic ray muons}

Experimental data on the muon flux in water \cite{Rogers1984}, as well as the data on the reaction cross-sections and the calculation results from \cite{Heisinger2002a, Heisinger2002b}, were used to calculate the rate of $\rm ^{14}C$ production by muons.

The production of the considered cosmogenic nuclide can result from inelastic interactions between high-energy muons and matter nuclei. Muons with energies close to the average muon flux energy mostly contribute to the nuclide production \cite{Heisinger2002a}. In ice nuclide is mainly produced by high-energy muons at depths of about ten meters of water equivalent. The average muon energy in matter is about 10 GeV at such depths \cite{Heisinger2002a}. According to our calculations, the ratio of omnidirectional fluxes of high-energy muons with energies higher than 5 GeV for periods of solar activity minima and maxima is 1.07 in the lower atmospheric layers.

At the end of its motion, a negative muon is trapped onto the outer shell by one ambient atom and descends to a lower atom energy level. Then, the muon either decays into an electron and neutrino or reacts with a nucleus. In the latter case, the nuclear reaction can result in the production of a cosmogenic radionuclide. We consider that muons, the energy of which is lower than the energy of the differential muon flux maximum, i.e., about 0.5 GeV according to \cite{Boezio2003}, are "stopping" muons. According to our calculations, the ratio of omnidirectional fluxes of low-energy negative muons with energies lower than 0.3 GeV during periods of solar activity minima and maxima is 1.2-1.3 for atmospheric altitudes not higher than 4 km.

Fluxes of cosmic ray muons in the atmosphere are less sensitive to a change in the solar activity level than those of nuclear-interacting component particles.

\subsection{Production of $\rm ^{14}C$ in Ice at Different Altitudes above Sea Level}

This subsection presents the calculation results of the integral radiocarbon production rate in ice by cosmic ray components $Q_{i}$ depending on the glacier height; $Q_{i} = \int_{0}^{\infty} P_{i}(z) dz$, where $P_{i}(z)$ is the radiocarbon production rate in ice by cosmic ray component $i$ at depth $z$. The value of the integral radiocarbon production rate determines the nuclide concentration in regions where polar ice is accumulated \cite{Nesterenok2010}.

The cosmogenic radionuclide production rate in ice is proportional to the omnidirectional integral flux of particles near the Earth$^\prime$s surface. According to the results presented in \cite{Dunai2000, Desilets2003}, the nuclide production rate by neutrons of cosmic rays in ice at height $h$ above sea level may be presented as

\begin{equation}
	Q_{n}^{h} = Q_{n}^{\rm s.l.} {\rm exp} \left(\frac{1034 - x}{\Lambda_{n}^{\rm atm}} \right)
	\label{q1}
\end{equation}

\noindent
where $Q_{n0}^{\rm h}$ is the nuclide integral production rate in ice at height $h$ above sea level; $Q_{n0}^{\rm s.l.}$ is such a rate at sea level; $x$ is the thickness of the atmosphere in $\rm g/cm^2$ above a specified level at height $h$, and $\Lambda_{n}^{\rm atm}$ is the attenuation length of the high-energy neutron integral omnidirectional flux at atmospheric altitudes varying from 0 to $h$. According to our modeling, the average attenuation length of the high-energy neutron omnidirectional flux (neutron energies higher than 10 MeV) is $\Lambda_{n}^{\rm atm} = 130 \pm 0.3$, $134 \pm 0.3$, and $138 \pm 0.4$ $\rm g/cm^2$ at the level of one standard error for minimal, moderate, and maximal solar activity, respectively, and for atmospheric altitudes no higher than 4 km. According to the results presented in \cite{Desilets2003} for high geomagnetic latitudes and altitudes reaching several kilometers, $\Lambda_{n}^{\rm atm}$ = 130 + 7 - 3 $\rm g/cm^2$, and the variation in the parameter during the 11-year solar cycle is about 4\% \cite{Raubenheimer1974}, which confirms the results achieved by us in the present work.

The muon flux attenuation length in the atmosphere is a function of the muon energy. A linear dependence is observed between the attenuation length of a vertical differential muon flux and particle momentum \cite{Boezio2000}. According to \cite{Desilets2003}, the attenuation length of an omnidirectional flux of negative muons with an energy lower than 0.3 GeV is 230-240 $\rm g/cm^2$ for high geomagnetic latitudes and altitudes reaching several kilometers. According to our calculations, the attenuation length of an omnidirectional flux of low-energy muons is about 250 $\rm g/cm^2$ for atmospheric altitudes not higher than 4 km. The nuclide production rate in reactions of capture of negative muons in ice at height $h$ above sea level is

\begin{equation*}
	Q_{\mu^{-}}^{h} = Q_{\mu^{-}}^{\rm s.l.} {\rm exp} \left(\frac{1034 - x}{\Lambda_{\mu^{-}}^{\rm atm}} \right)
\end{equation*}

\noindent
where $Q_{\mu^{-}}^{h}$, $Q_{\mu^{-}}^{\rm s.l.}$ and $x$ have the same meaning as in expression (\ref{q1}) and $\Lambda_{\mu^{-}}^{\rm atm}$ is the attenuation length of the omnidirectional flux of "stopping" muons; in these calculations, $\Lambda_{\mu^{-}}^{\rm atm}$ = 240 $\rm g/cm^2$.

The nuclide production rate by high-energy muons in ice at height $h$ above sea level is

\begin{equation*}
	Q_{\mu f}^{h} = Q_{\mu f}^{\rm s.l.} {\rm exp} \left(\frac{1034 - x}{\Lambda_{\mu f}^{\rm atm}} \right)
\end{equation*}

\noindent
where $\Lambda_{\mu f}^{\rm atm}$ is the attenuation length of the omnidirectional flux of high-energy muons in the atmosphere. We assume that the $\Lambda_{\mu f}^{\rm atm}$ value is equal to the attenuation length of a muon flux with energies of 5-10 GeV in the atmosphere. According to the results presented in \cite{Boezio2000} and our calculations, $\Lambda_{\mu f}^{\rm atm} \approx 1000 \; \rm g/cm^2$. Figure 3 presents the dependences of the rate of radiocarbon production by different cosmic ray components in ice on the glacier height at high geomagnetic latitudes and moderate solar activity. The total integral rate of radiocarbon production is also presented here.

\begin{figure}[t]
	\centering
	\includegraphics[width = 0.65\textwidth]{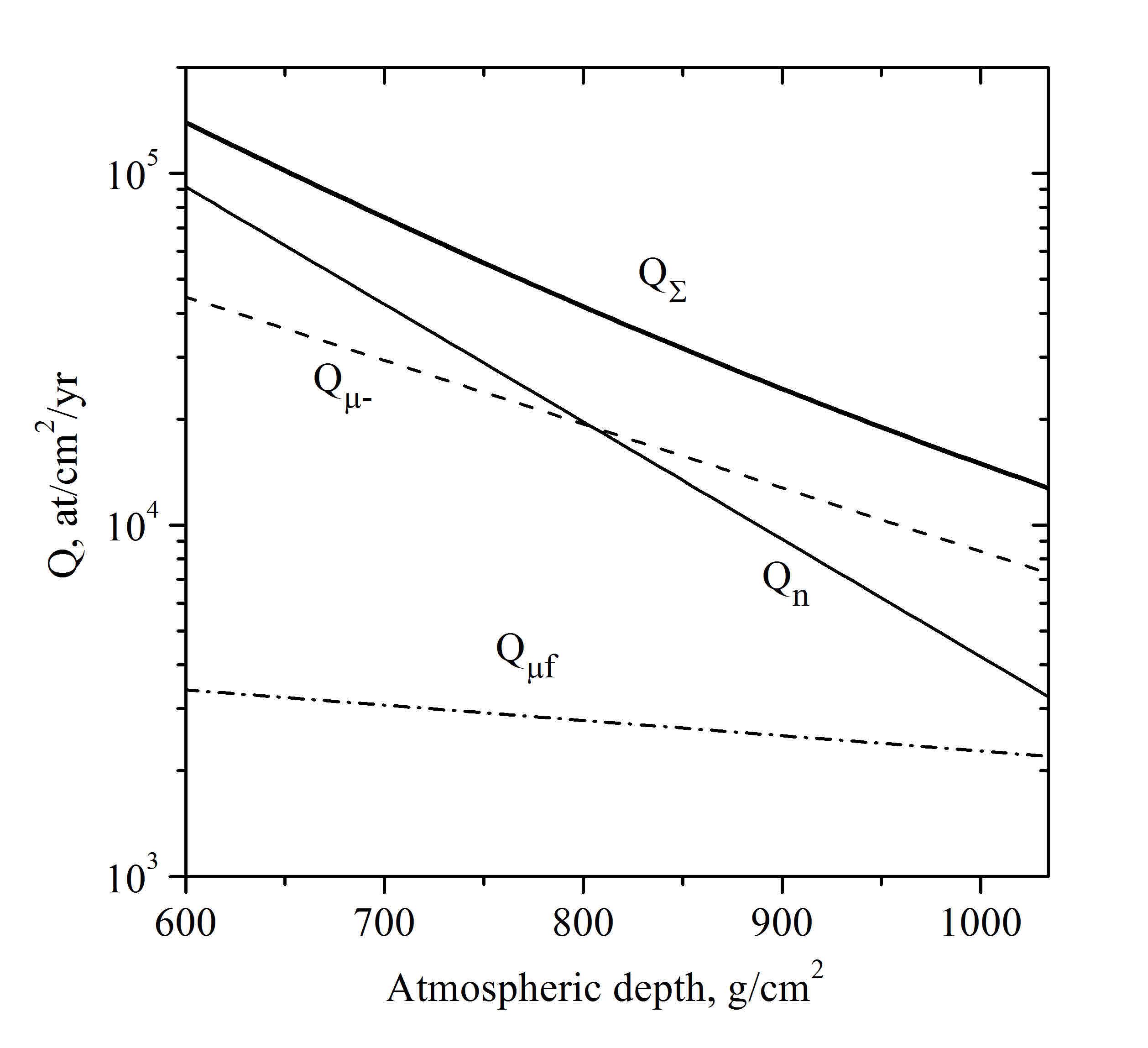}
		\caption{Integral rates of $\rm ^{14}C$ production in ice by cosmic ray components in reactions caused by cosmic ray neutrons $Q_{n}$ (solid thin curve) and high-energy muons $Q_{\mu f}$ (dot-and-dash curve) and in reactions of trapping negative muons $Q_{\mu^{-}}$ (dashed curve); total production rate $Q_{\Sigma}$ is represented by a thick solid curve.}
		\label{fig3}
\end{figure}

\section{Conclusions}
\noindent
We presented here the simulation results of the propagation of cosmic rays in the Earth$^\prime$s atmosphere for solar activity minima and maxima. We considered data on solar activity in the current epoch. According to the calculations, the ratio of fluxes of high-energy neutrons at minimal and maximal solar activity can reach 1.6 for the near-Earth atmospheric layers. At the same time, this ratio is less than 1.1 for high-energy muons. The contribution of the cosmic ray neutrons to the total integral radiocarbon production rate increases with increasing glacier height and is predominant for heights of more than 2 km. The $\rm ^{14}C$ production rate in ice at high altitudes and high geomagnetic latitudes depends substantially on the solar activity level.

\section*{Acknoledgements}
\noindent
This work was supported by the Ministry of Education and Science of the Russian Federation (contract no. 11.G34.31.0001 with SPbGPU and G.G. Pavlov, Leading Scientist).

\end{document}